\documentclass[a4paper,twoside]{article}

\usepackage{epsfig}
\usepackage{subcaption}
\usepackage{calc}
\usepackage{amssymb}
\usepackage{amstext}
\usepackage{amsmath}
\usepackage{amsthm}
\usepackage{multicol}
\usepackage{pslatex}
\usepackage{apalike}
\usepackage{algorithm2e}
\usepackage[bottom]{footmisc}
\usepackage{SCITEPRESS}
\usepackage{tikz}
\usetikzlibrary{quantikz2}
\usepackage{adjustbox}
\usepackage[hidelinks]{hyperref}
\usepackage[capitalize]{cleveref}
\creflabelformat{equation}{#2#1#3}

\begin{document}

\title{Multi-Agent Quantum Reinforcement Learning using Evolutionary Optimization}

\author{\authorname{Michael Kölle\sup{1}, Felix Topp\sup{1}, Thomy Phan\sup{2}, Philipp Altmann\sup{1}, Jonas Nüßlein\sup{1} and Claudia Linnhoff-Popien\sup{1}}
\affiliation{\sup{1}Institute of Informatics, LMU Munich, Munich, Germany}
\affiliation{\sup{2}Thomas Lord Department of Computer Science, University of Southern California, Los Angeles, USA}
\email{michael.koelle@ifi.lmu.de}
}


\abstract{
Multi-Agent Reinforcement Learning is becoming increasingly more important in times of autonomous driving and other smart industrial applications. Simultaneously a promising new approach to Reinforcement Learning arises using the inherent properties of quantum mechanics, reducing the trainable parameters of a model significantly. However, gradient-based Multi-Agent Quantum Reinforcement Learning methods often have to struggle with barren plateaus, holding them back from matching the performance of classical approaches. While gradient free Quantum Reinforcement Learning methods may alleviate some of these challenges, they too are not immune to the difficulties posed by barren plateaus. We build upon an existing approach for gradient free Quantum Reinforcement Learning and propose three genetic variations with Variational Quantum Circuits for Multi-Agent Reinforcement Learning using evolutionary optimization. We evaluate our genetic variations in the Coin Game environment and also compare them to classical approaches. We showed that our Variational Quantum Circuit approaches perform significantly better compared to a neural network with a similar amount of trainable parameters. Compared to the larger neural network, our approaches archive similar results using $97.88\%$ less parameters.
}

\onecolumn \maketitle \normalsize \setcounter{footnote}{0} \vfill

\section[Introduction]{\uppercase{Introduction}} \label{sec:introduction}

Artificial intelligence (AI) continues to advance, offering innovative solutions across various domains. Key applications include autonomous driving \cite{DBLP:journals/corr/Shalev-ShwartzS16a}, the internet of things \cite{9001216}, and smart grids \cite{5589633}. Central to these applications is the use of Multi-Agent Systems (MAS). These agents, though designed to act in their own interest, can be guided to work together using Multi-Agent Reinforcement Learning (MARL). Notably, MARL has proven effective, especially in resolving social dilemmas \cite{DBLP:journals/corr/LeiboZLMG17}.

Reinforcement Learning (RL) itself has made impressive strides, outperforming humans in areas like video games \cite{DBLP:journals/corr/abs-2003-13350,DBLP:journals/corr/abs-1911-08265}. Alongside this, quantum technologies are emerging, suggesting faster problem-solving and more efficient training in RL \cite{Harrow_2017}. However, Quantum Reinforcement Learning (QRL) has it's challenges, such as instabilities and vanishing gradients \cite{instabilities,chen2022variational}. To address these, researchers have turned to evolutionary optimization methods, as proposed by \cite{chen2022variational}, which have shown promising results. With the rising prominence of MARL, combining it with quantum techniques has become a research focal point, leading to the development of Multi-Agent Quantum Reinforcement Learning (MAQRL). 
In this work, each agent is represented as a Variational Quantum Circuits (VQC). We employ a evolutionary algorithm to optimize the parameters of the circuit. We evaluate different generational evolution strategies and conduct a small scale hyperparameter search for key parameters of the VQC. Our aim is to evaluate MAQRL's capabilities and compare it to traditional RL methods, using the Coin Game as a benchmark.

In this study, we model each agent using Variational Quantum Circuits (VQC), a promising and adaptable representation in the quantum domain. The inherent flexibility of VQCs allows for the encoding of complex information, making them suitable for representing agent behaviors in diverse environments. To fine-tune these quantum circuits and ensure their optimal performance, we harness the power of an evolutionary algorithm. This algorithm iteratively optimizes the parameters of the VQC, guiding the circuit towards improved decision-making and interactions. While evolutionary algorithms have been traditionally employed in classical domains, their application in the quantum realm offers exciting prospects for efficiently navigating the vast parameter space of VQCs. As part of our experiements, we systematically evaluate multiple generational evolution strategies. By comparing their effectiveness, we aim to identify which strategies most beneficially influence the learning trajectories of the VQCs. Furthermore, recognizing the significance of the VQC's parameters in determining its behavior and effectiveness, we undertake a small scale hyperparameter search. This search is dedicated to fine-tuning key parameters, ensuring the VQC. Central to our research objectives is the evaluation of MAQRL and its potential contributions to the field. We are particularly interested in benchmarking MAQRL against established RL techniques to learn its advantages and areas of improvement. For a robust and fair assessment, we have chosen the Coin Game, a well-regarded environment in multi-agent research, as our testing ground. In summary our contributions are:
\begin{enumerate}
    \item Introducting evolutionary optimization in a quantum multi-agent reinforcement learning setting.
    \item Assessing the impact of three different generational evolution strategies and variational layer counts.
    \item Direct comparison to classical approaches with different parameter counts.
\end{enumerate}

\noindent We start in \cref{sec:preliminaries} by explaining the basics of MARL and Evolutionary Optimization. We also give a short introduction to Quantum Computing and VQCs, and mention related studies (\cref{sec:related-work}). After outlining our methodology (\cref{sec:approach}) and experimental setup (\cref{sec:experimental-setup}), we share the results and implications of our experiments in \cref{sec:results}. We end with a summary and thoughts on next steps for research (\cref{sec:conclusion}). All code and experiments can be found here\footnote{https://github.com/michaelkoelle/qmarl-evo}. 
\section[Preliminaries]{\uppercase{Preliminaries}} \label{sec:preliminaries}
\subsection{Multi-Agent Setting}
We focus on \emph{Markov games} $M = \langle \mathcal{D},\mathcal{S},\mathcal{A},\mathcal{P},\mathcal{R} \rangle$, where $\mathcal{D} = \{1,...,N\}$ is a set of agents $i$, $\mathcal{S}$ is a set of states $s_{t}$ at time step $t$, $\mathcal{A} = \langle \mathcal{A}_{1}, ..., \mathcal{A}_{N} \rangle$ is the set of joint actions $a_{t} = \langle a_{t,i} \rangle_{i \in \mathcal{D}}$, $\mathcal{P}(s_{t+1}|s_{t}, a_{t})$ is the transition probability, and $\langle r_{t,1}, ..., r_{t,N} \rangle = \mathcal{R}(s_{t},a_{t}) \in \mathbb{R}$ is the joint reward. $\pi_{i}(a_{t,i}|s_{t})$ is the action selection probability represented by the individual \emph{policy} of agent $i$.

Policy $\pi_{i}$ is usually evaluated with a \emph{value function} $V_{i}^{\pi}(s_{t}) = \mathbb{E}_{\pi}[G_{t,i}|s_{t}]$ for all $s_{t} \in \mathcal{S}$, where $G_{t,i} = \sum_{k=0}^{\infty} \gamma^{k} r_{t+k,i}$ is the individual and discounted \emph{return} of agent $i \in \mathcal{D}$ with discount factor $\gamma \in [0,1)$ and $\pi = \langle \pi_{1}, ..., \pi_{N} \rangle$ is the \emph{joint policy} of the MAS. The goal of agent $i$ is to find a \emph{best response} $\pi_{i}^{*}$ with $V_{i}^{*} = max_{\pi_{i}}V_{i}^{\langle \pi_{i}, \pi_{-i} \rangle}$ for all $s_{t} \in \mathcal{S}$, where $\pi_{-i}$ is the joint policy \emph{without} agent $i$.

We define the \emph{efficiency} of a MAS or \emph{utilitarian metric (U)} by the sum of all individual rewards until time step $T$:
\begin{equation}\label{eq:fitness}
U = \sum_{i \in \mathcal{D}}R_{i}
\end{equation}
where $R_{i} =  \sum_{t=0}^{T-1} r_{t,i}$ is the \emph{undiscounted return} or \emph{sum of rewards} of agent $i$ starting from start state $s_{0}$.

\subsection{Multi-Agent Reinforcement Learning}

We focus on independent learning, where each agent $i$ optimizes its individual policy $\pi_{i}$ based on individual information like $a_{t,i}$ and $r_{t,i}$ using \emph{RL} techniques, e.g., evolutionary optimization as explained in Section \emph{Evolutionary Optimization}. Independent learning introduces non-stationarity due to simultaneously adapting agents which continuously changes the environment dynamics from an agent's perspective (\cite{littman1994markov,laurent2011world,hernandez2017survey}), which can cause the adoption of overly greedy and exploitative policies which defect from any cooperative behavior (\cite{leibo2017multi,foerster2018learning}).

\subsection{Evolutionary Optimization}
\label{sec:evolutionary_ptimization}
Inspired by the process of natural selection, evolutionary optimization have been shown to find optimal solutions to complex problems, where traditional methods may not be efficient \cite{7955308}. 
They employ a population of individuals, randomly generated, each with its own set of parameters. 
These individuals are evaluated based on a fitness function that measures how well their parameters perform on the given problem. 
The fittest individuals are then selected for reproduction, where their parameters are recombined and mutated to form a new population of individuals for the next generation. \cite{eiben2015introduction}

Evolutionary optimization approaches like genetic algorithms \cite{holland1991artificial} have been used successfully in a variety of fields, including the optimization of neural networks, or in interactive recommendation tasks \cite{ding2011optimizing,gabor2019benchmarking}. 
Furthermore these methods have been used to solve a wide range of problems, from designing quantum circuit architectures to optimizing complex real-world designs \cite{lukac2002evolving,caldas2002design}.

\subsection{Quantum Computing}
\label{sec:qc}
Quantum computing is a emerging field of computer science that uses the principles of quantum mechanics to process information. Similar to classical computers, which store and process data as bits, quantum computers use quantum bits, or qubits, which can reside in multiple states at once \cite{yanofsky2008quantum}. This property is called \emph{superposition}. A state $|\psi\rangle$ of a qubit can generally be expressed as a linear combination of $|0\rangle$ and $|1\rangle$
\begin{equation}
|\psi\rangle = \alpha|0\rangle + \beta|1\rangle,
\label{qubit}
\end{equation}
where $\alpha$ and $\beta$ are complex coefficients that satisfy the equation
\begin{equation}
|\alpha|^2 + |\beta|^2 = 1.
\end{equation}
When a qubit in the state of $\alpha \vert 0 \rangle + \beta \vert 1 \rangle$ is measured, its superposition collapses into one of its possible states, either $|$0⟩ or $|$1⟩, with probabilities determined by the coefficients $|\alpha|^2$ and $|\beta|^2$ respectively \cite{mcmahon2007quantum}. The quantum system transitions from a superposition of states to an actual classical state where the observable's value is precisely known \cite{nielsen_chuang_2010}. Multiple qubits can be bound together via \emph{entanglement} to archive strong correlations between them.

\subsection{Variational Quantum Circuits}
\label{sec:vqc}
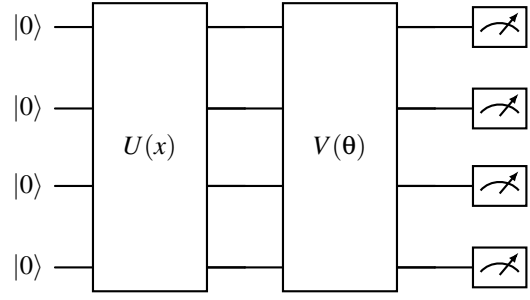
\begin{figure}[!htbp]
    \begin{center}
        \begin{quantikz}
        \lstick{\ket{0}} & \gate[wires=4][1.5cm]{U(x)} & \qw & \gate[wires=4][1.5cm]{V(\theta)} & \qw & \meter{} \\
        \lstick{\ket{0}} && \qw && \qw & \meter{} \\
        \lstick{\ket{0}} && \qw && \qw & \meter{} \\
        \lstick{\ket{0}} && \qw && \qw & \meter{}
        \end{quantikz}
    \end{center}
    \caption{Structure of a Variational Quantum Circuit}
    \label{fig:vqc}
\end{figure}

VQC, also known as parameterized quantum circuits, are quantum algorithms that act as function approximators and are trained using a classical optimization process. They are commonly used as a drop-in replacement for Neural Networks for Deep RL \cite{chen2022variational,schuld2020circuit,DBLP:journals/corr/abs-1907-00397,skolik2021layerwise,https://doi.org/10.48550/arxiv.2210.14876}. A VQC is made up of three stages, as can be seen in \cref{fig:vqc}. First, the classical input is embedded into a quantum state in the  \emph{State Preperation} stage $U(x)$ using superposition. In the \emph{Variational Layers} stage $V(\theta)$, qubits are then entangled and parameterized for training. Finally, in the \emph{Measurement} stage, the output of the circuit is measured repeatedly to get the expectation value of each qubit.

\noindent \textbf{State Preperation}: In this work, we use Amplitude Embedding \cite{mottonen2004transformation} to encode classical data into a quantum state. As the name suggests, the features are embedded into the amplitudes of the qubits. Using superposition, we can embed $2^n$ features into $n$ qubits. For Example, if we want to embed feature vector $x \in \mathbb{R}^3$ in to a 2 qubit quantum state $\ket{\psi} = \alpha\ket{00} + \beta\ket{01} + \gamma\ket{10} + \delta\ket{11}$ such that $\lvert \alpha\rvert^2 + \lvert \beta\rvert^2 + \lvert \gamma\rvert^2 + \lvert \delta\rvert^2 = 1$, we first pad our feature vector so that it matches $2^n$ features where $n$ is the number of qubits used. Next, we normalize the padded feature vector $y$ such that $\sum_{k=0}^{2^n - 1} \frac{y_k}{\lvert\lvert y\rvert\rvert} = 1$. Lastly, we use the state preperation by Mottonen et al. \cite{mottonen2004transformation} to embed the padded and normalized feature vector into the amplitudes of the qubit state.
\\\\
\textbf{Variational Layers}
The second part of the circuit, referred to as Variational Layers, is made up of repeated single qubit rotations and entanglers (\cref{fig:usedvqc}, everything within the dashed blue area is repeated $L$ times, where $L$ is the layer count). We use a layer architecture inspired by the circuit-centric classifier design \cite{schuld2020circuit} in particular. All of the circuits presented in this paper employ three single qubit rotation gates and CNOT gates as entanglers.\\

\begin{adjustbox}{width=\linewidth}
\begin{quantikz}
     \qw& \ctrl{1} & \qw      & \targ{} & \gate{RZ(\theta_0^0)} & \gate{RY(\theta_0^1)}& \gate{RZ(\theta_0^2)} &  \qw   \\
     \qw& \targ{}  & \ctrl{1} & \qw     & \gate{RZ(\theta_1^0)} & \gate{RY(\theta_1^1)}& \gate{RZ(\theta_1^2)} &  \qw    \\
     \qw& \qw      & \targ{}  & \ctrl{-2}&\gate{RZ(\theta_2^0)} & \gate{RY(\theta_2^1)}& \gate{RZ(\theta_2^2)} &  \qw 
\end{quantikz}
\end{adjustbox}
\\\\\\
$\theta i^j$ denotes a trainable parameter in the circuit above, where $i$ represents the qubit index and $j \in \{0,1,2\}$ the index of the single qubit rotation gate. For simplicity, we omitted the index $l$, which denotes the current layer in the circuit. The target bit of the CNOT gate in each layer is given by$(i + l)\ \text{mod}\ n$.
\\\\
\textbf{Measurement}
The expectation value is measured in the computational basis ($z$) of the first $k$ qubits, where $k$ is the dimension of the agents' actions space. Each measured expectation value is then given a bias. The biases are also included in the VQC parameters and are updated accordingly.

\section[Related Work]{\uppercase{Related Work}} \label{sec:related-work}

QRL is progressively gaining traction, emanating from the intersection of RL and the emerging field of QC \cite{chen2022variational,kwak2021introduction}. This chapter delves into diverse applications and theoretical concepts within QRL that yield advancements in parameter reduction, expedited computation times, and addressing intricate problems. 

Initially, we focus on a method by Chen et al., wherein parameters are refined using an evolutionary approach \cite{chen2022variational}, forming the foundation upon which the current work is built. Evolutionary algorithms have established their efficacy within traditional RL \cite{DBLP:journals/corr/abs-1712-06567} and have demonstrated substantial value for Deep Reinforcement Learning (DRL). The methodology employed within this research integrates DRL strategies, substituting Neural Networks with Variational Quantum Circuits as agents. Chen et al. demonstrated, in a discrete environment, that VQCs can efficiently approximate Q-value functions in Deep Q-Learning (DQL), presenting a quantum perspective to RL, while notably reducing parameter requirements in comparison to classical RL. Differing from Chen, the approach presented in this work incorporates recombination and extends the gradient free method to the domain of MARL. 

An alternative route to Quantum Multi-Agent Reinforcement Learning is detailed in \cite{neumann2020multi,maqbmrl-22} which harnesses Quantum Boltzmann Machines (QBM). The strategy originates from an extant methodology where QBM outperforms classical DRL in convergence speed, measured in the number of requisite time steps, utilizing Q-value approximation. The outcomes hint at enhanced stability in learning, and the agents attaining optimal strategies in Grid domains, surmounting the complexity of the original approach. This method may serve as a foundational approach in Grid domains and in other RL domains with superior complexity. Resemblances to the method employed herein lie in the utilization of Q-values and grid domains as testing environments. 

Moreover, \cite{yun2022quantum} explores the application of VQCs for QRL and the progression of this concept to QMARL, bearing similarity to the methodology delineated in this thesis. The limited number of parameters in QRL has demonstrated superior outcomes compared to classical computing. In order to navigate the challenges of extending this to QMARL in the Noisy Intermediate-Scale Quantum era and the non-stationary attributes of classical MARL, a strategy of centralized learning and decentralized execution is enacted. This approach achieves an overall superior reward in the environments tested, compared to classical methodologies, with disparities arising in the employment of evolutionary algorithms and the architecture of the VQCs.

\section[Approach]{\uppercase{Approach}} \label{sec:approach}


Inspired by to \cite{chen2022variational}, we propose to employ an evolutionary approach to optimize a $\theta$ parameterized agent.
We however consider the more general Multi-Agent setup introduced above. 
Thus, we aim to optimize the utilitarian metric $U$ (cf. Eq.~\eqref{eq:fitness}).
To maximize this fitness function we use a population $\mathcal{P}$ consisting of $\eta$ random initialized agents, parameterized by $\theta\in[-\pi,\pi]$ to match the quantum circuits' parameter space. 


In contrast to previous work, we use VQC rather than neural networks to approximate the value of the agent's  actions.
This should mainly demonstrate the improved parameter efficiency, as previously denoted, even applied to complex learning tasks.
A VQC consists primarily of three components: the input embedding, the repeated variational layers, and the measurement. \cref{fig:vqc} depicts the VQC we employ.

\begin{figure}
    \centering
    \begin{adjustbox}{width=\linewidth}
    \begin{quantikz}
    \lstick[wires=6]{$\ket{0}$} & \gate[wires=6]{U(x)} & \ctrl{1}\gategroup[6,steps= 8,style={dashed,
    rounded corners,fill=blue!20, inner xsep=2pt},
    background]{{\sc Variational Layer}} & \qw & \qw & \qw & \qw &  \targ{} & \qw & \gate{R(\alpha _1, \beta _1, \gamma _1)}& \meter{} \\
    & \qw &\targ{} & \ctrl{1} & \qw &  \qw & \qw & \qw & \qw &\gate{R(\alpha _2, \beta _2, \gamma _2)} & \meter{}\\
    & \qw & \qw & \targ{} & \ctrl{1} & \qw & \qw & \qw & \qw & \gate{R(\alpha _3, \beta _3, \gamma _3)} & \meter{}\\
    & \qw & \qw & \qw & \targ{} & \ctrl{1} \qw & \qw & \qw & \qw &\gate{R(\alpha _4, \beta _4, \gamma _4)}&\meter{}\\
    & \qw & \qw & \qw & \qw & \targ{} & \ctrl{1} & \qw & \qw & \gate{R(\alpha _5, \beta _5, \gamma _5)} \\
    & \qw & \qw & \qw & \qw & \qw & \targ{} & \ctrl{-5}& \qw & \gate{R(\alpha _6, \beta _6, \gamma _6)}
    \end{quantikz}
    \end{adjustbox}
    \caption{Variational Quantum Circuit}
    \label{fig:usedvqc}
\end{figure}
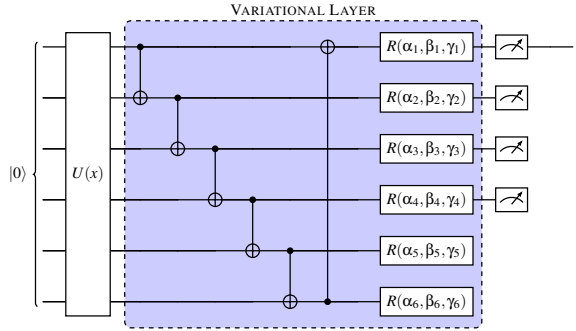

To convert the classical data into a quantum state, we use Amplitude Embeddings, represented by $U(x)$ in \cref{fig:usedvqc}.
Caused by the high dimensionality of most state spaces, Amplitude Embeddings are currently the only viable embedding strategy that allow for embedding the whole state information, being able to embed $2^{n_q}$ states in $n_q$ qubits.

The second part of the VQCs consists of variational layers that are variably repeated. 
Each iteration increases the number of $\alpha_i,\beta _i,\gamma _i$ parameters that are defined by $\theta$ and make up each individual to be optimized.
Per layer, there are $n_\theta = n_q * 3$ parameters, where $n_q$ is the number of qubits. 
Furthermore, all rotations are performed sequentially as $R_Z(\alpha _i), R_Y(\beta _i)$ and $R_Z(\gamma _i)$.
In addition to the parameterized rotations, each variational layer is composed of adjacent CNOTS to entangle all qubits. 
After $n_l$ repetitions of the variational layer, the predicted values of the individual actions are determined by measuring the first $n_a$ qubits, where $n_a$ is the number of actions. 
This Z-axis measurement is used to determine the Q-value of the corresponding action. 
An agent chooses the action with the greatest expected value.

The proposed evolutionary algorithm training procedure to optimize these individuals to maximize the utilitarian metric $U$ is demonstrated in \cref{alg:evo}. 

\begin{algorithm}
\DontPrintSemicolon  

\caption{Evolutionary optimization algorithm}\label{alg:evo}

\KwData{Population Size $\eta$, Number of Generations $\mu$, Evaluation Steps $\kappa$, Truncation Selection $\tau$, Mutation Power $\sigma$, and Number of Agents $N$}
\KwResult{Population $\mathcal{P}$ of optimized agents}

$\mathcal{P}_0 \gets$ Initialize population $\eta$ with random $\theta$\;

\For{$g \in \{0,1,...,\mu\}$}{
    \For{$i \in \{0,1,...,\eta\}$}{
        Reset testing environment\;
        Score $S_{t,i} \gets 0$\;
        \For{$t \in \{0,1,...,\kappa\}$}{
            Use policy of agent $i$ for all agents in env\; 
            Select action $a_t \gets \text{argmax}\ VQC_\theta(s_t)$\;
            Execute environment step with action $a_t$\;
            Observe reward $r_t$ and next state $s_{t+1}$\;
            $S_{t,i} \gets S_{t,i} + r_t$\;
        }
    }
    $\lambda \gets$ Select top $\tau$ agents based on $S_{t,i}$\;
    Keep top agent based on $S_{t,i}$\;
    Recombine $\eta - 1$ new agents out of $\lambda$\;
    Mutate $\eta - 1$ generated agents\;
    $\mathcal{P}_{g+1} \gets$ $\eta - 1$ generated agents + top agent\;
}

\end{algorithm}

For each generation, first, the fitness of each individual $i$ is evaluated by performing $\kappa$ steps in the environment.
Building upon this fitness, the best $\tau$ agents are selected to develop a new generation. 
In addition, we employ the so-called elite agent, the agent with the highest fitness, that is excluded from the following mutation procedure. 

To form the next generation, mutation and recombination possibilities are combined to generate a new population.
First, new individuals are formed by recombining the $\kappa$ best agents of the current generation using crossover.
The new offspring is produced by randomly selecting two parents and crossing their parameters at a randomly selected index.
Furthermore, mutation is applied to generate new agents by modifying the parameters $\theta$ of the current generation of the best $\tau$ agents:
\begin{equation}
    \theta = \theta + \sigma * \epsilon
\end{equation}
The agents with the highest fitness values $T$ are the parents of the upcoming generation. For the mutation, the parameters $\theta$ are modified as seen in the equation with the mutation power $\sigma$ the Gaussian noise $\epsilon \sim \mathcal{N(0,1)}$. 
Consequently, all $\theta _i$ parameters undergo a minor mutation, and new agents, or children, are generated.
Finally, the unaltered elite agent is added to the child population.



\section[Experimental Setup]{\uppercase{Experimental Setup}} \label{sec:experimental-setup}
\subsection{Coin Game Environment}
\label{sec:Coin}
\begin{figure}[ht]
    \centering
    \includegraphics[width=0.3\textwidth]{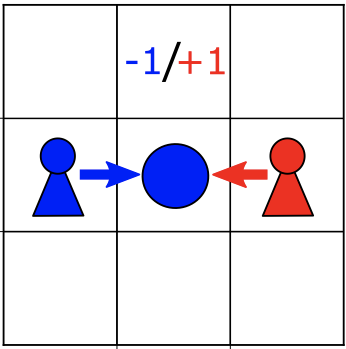}
    \caption{Example State of the Coin Game by \cite{phanAAMAS22}.}
    \label{fig:CoinGame}
\end{figure}

As of today, we are in the Noisy Intermediate-Scale Quantum era of quantum computing, where we can simulate only a small number of qubits \cite{Preskill2018}. This heavily restricts the amount of data we can embed into the quantum circuit. Therefore, we are limited in our choices for the evaluation environment. We chose the Coin Game environment in its $3\times3$ gridworld version due to its relatively small observation space. The Coin Game, which was created by \cite{lerer2017maintaining}, is a well-known sequential game for assessing RL strategies. Both the Red and Blue agents in the environment are tasked with collecting coins. Beside the agents there is a single coin placed in the grid that corresponds with one of the agents colors. \cref{fig:CoinGame} depicts a exemplary state within the Coin Game.

When an agent is in the same position as a coin, it is deemed collected. After a coin is gathered, a new coin is generated at a random location, which cannot be a place occupied by an agent, and is again either red or blue. A game of the Coin Game is limited to 50 steps 25 per agent, and the objective is to maximize the agents' rewards. The Coin Game can be played in both a competitive and cooperative setting. To make the game cooperative, the reward for collecting a coin is increased by +1 for the agent who collects the coin. Moreover, the second agent's reward is reduced by -2 if the first agent obtains a coin of his color. If we now consider the agents' total reward, there is a common reward of +1 for collecting an own coin and -1 for collecting an opposing coin. This causes the agents to be trained to gather their own coins and leave coins, which are not the agent's color, for the other agent to collect. If both agents performed random behaviors, the expected reward should be zero. As a result of these rewards, the coin game is a zero-sum game.

Each cell of the $3\times3$ gridworld can contain either agent 1, agent 2, a red or blue coin. Empty grids need not be included in the observation, as movement can occur on them without consequence at any time. An agent may select from four possible actions, each of which is only possible if the ensuing movement does not lead outside the $3\times3$ gridworld. The numerical actions range from 0 to 3. Action 0 indicates a step to the north, action 1 a step to the south, action 2 a step to the west, and action 3 a step to the east. To prevent the VQC from choosing illegal actions, the expected values are normalized to the interval $[0,1]$ and masked with environmental regulations. 

\subsection{Baselines}
\label{sec:NN}
VQC can be viewed as an alternative to the employment of classical neural networks as agents. In the framework of our methodology, we employ neural networks as agents, as they can be considered general approximators. We employ a 2-layer basic neural network for this purpose. The inputs in the form of observations are mapped to a variable number x of hidden units in the first layer. The second layer relates the number of our activities to x hidden units. Hence, we obtain the individual Q-values for each action, just as we did with the VQC. Moreover, the Q-values are multiplied by the action mask to ensure that no illegal action can be selected. Here, we limit ourselves to this neural network with variable numbers of hidden units. Similar to the VQC, there are a great number of ways to alter the network and hence alter the results.

\subsection{Metrics}
We evaluate our experiments in the Coin Game environment using three metrics: Score, Total Coin Rate and Own Coin Rate. In our work, the agents are solely playing against themselves, to easily evaluate the agents' performance. The first metric, Score $S_n$ consists of the undiscounted individual rewards $r_{t,i}$ until timestep $T \in \{0..49\}$ accumulated over all agents

\begin{equation}\label{eq:score}
 S_n = \sum_{i \in \{0, 1\}} \sum_{t=0}^{T-1} r_{t,i}
\end{equation}

\noindent with agent $i$ and generation $n \in \{0..99\}$ averaged over five seeds. This is a good overall indicator of the agents performance in the Coin Game environment. The next two metrics should provide insight into how the score is reached. The total coins collected metric $TC_n$ is the sum of all collected coins $c_{t,i}$ by all agents until timestep $T \in \{0..49\}$

\begin{equation}\label{eq:total_coin}
 TC_n = \sum_{i \in \{0, 1\}} \sum_{t=0}^{T-1} c_{t,i}
\end{equation}

\noindent with agent $i$ and generation $n \in \{0..99\}$ averaged over five seeds. The own coins collected metric $OC_n$ is the sum of all collected coins that corresponds to the agents own color $o_{t,i}$ until timestep $T \in \{0..49\}$ accumulated over all agents

\begin{equation}\label{eq:own_coin}
 OC_n = \sum_{i \in \{0, 1\}} \sum_{t=0}^{T-1} o_{t,i}
\end{equation}

\noindent with agent $i$ and generation $n \in \{0..99\}$ averaged over five seeds. Comparing the latter two metrics, we can get a greater insight how much cooperation is archived, with the own coin rate $OCR$:

\begin{equation}\label{eq:own_coin_rate}
 OCR_n = \sum_{i \in \{0, 1\}} \sum_{t=0}^{T-1} \frac{o_{t,i}}{c_{t,i}}
\end{equation}

\subsection{Training and Hyperparameters}
\label{sec:hyparams}
For our experiments in the Coin Game environment, we train the agents for $\mu = 100$ generations with a population size of $\eta = 250$, pairing the agents against themselves to play a game of 50 steps 25 per agent. After a brief preliminary study we set the mutation power to $\sigma = 0.01$. We select the top $\tau = 5$ agents for regenerating the following population. The VQC has a Variational Layer count of $4$ and $n_q=6$ qubits to embed the 36 features of the coin game, resulting in a parameter count of $76$. Each experiment is conducted with five different seeds $\in {0..4}$ to provide a more accurate indication of performance. Due to current quantum hardware limitations we use the Pennylane DefaultQubit simulator for all VQC executions. All runs were executed on nodes with Intel(R) Core(TM) i5-4570 CPU @ 3.20GHz.
\section[Results]{\uppercase{Results}} \label{sec:results}
In this section, we present the results of our experiments in the Coin Game environment. We tested our approach with recombination and mutation combined, as well as mutation only. Furthermore, we tested two classical neural networks with two hidden layers, one with a hidden layer size of $64\times64$ and one with $3\times4$ respectively. The latter configuration closely matches the amount of parameters that the VQCs approaches use, to get better insights on the model-size/performance ratio. Finally, we ran tests with agents that take random actions at every step which forms our random baseline. The number of parameters is listed alongside each approach.

\begin{figure}[ht]
    \centering
    \includegraphics[width=\linewidth]{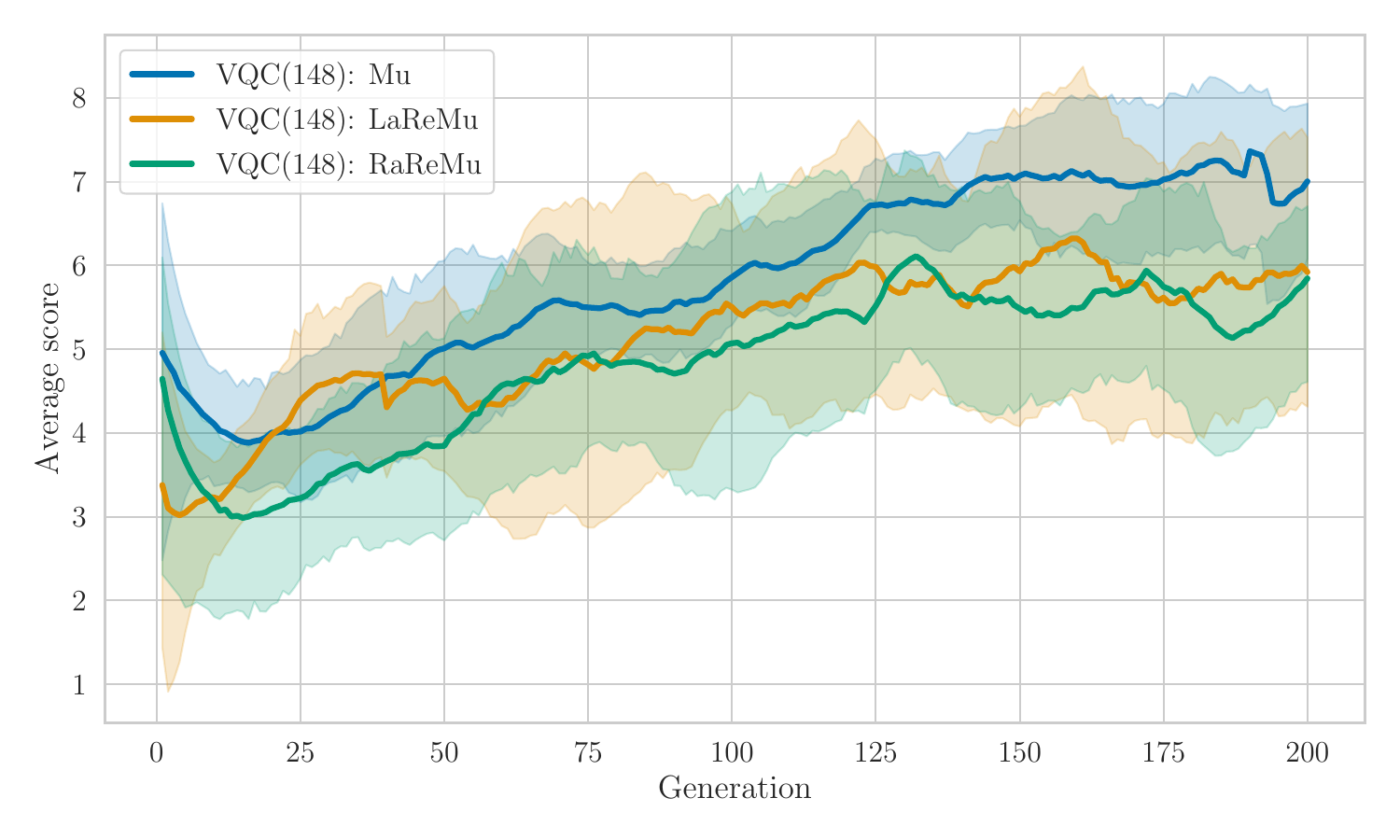}
    \caption{Average Score over the entire population. Each individual has completed 50 steps in the Coin Game environment each generation.}
    \label{fig:MethodsAvgScore}
\end{figure}

\begin{figure*}
     \centering
     \subfloat[][Total coins collected]{\includegraphics[width=0.32\textwidth]{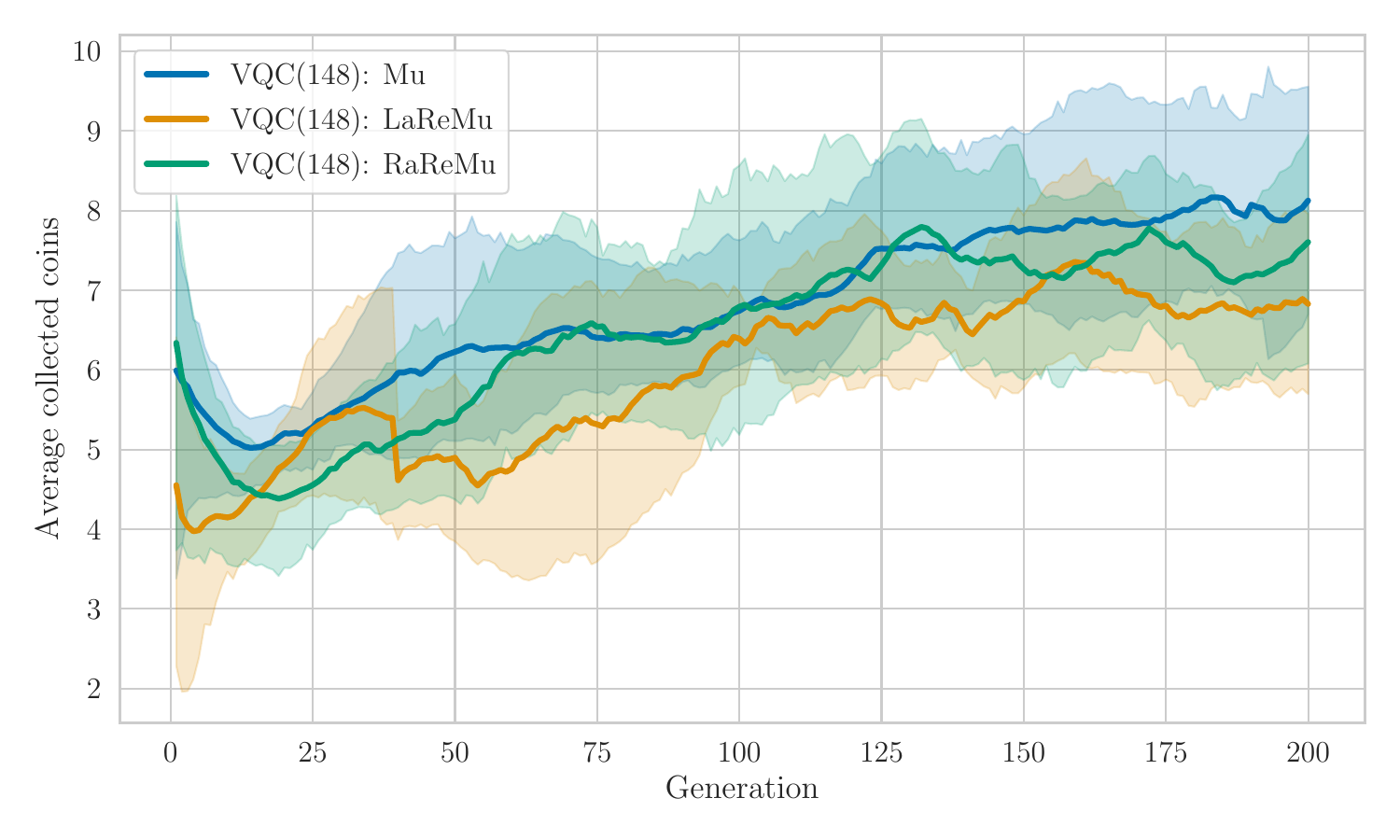}\label{fig:MethodsAvgCoins}}
     \subfloat[][Own coins collected]{\includegraphics[width=0.32\textwidth]{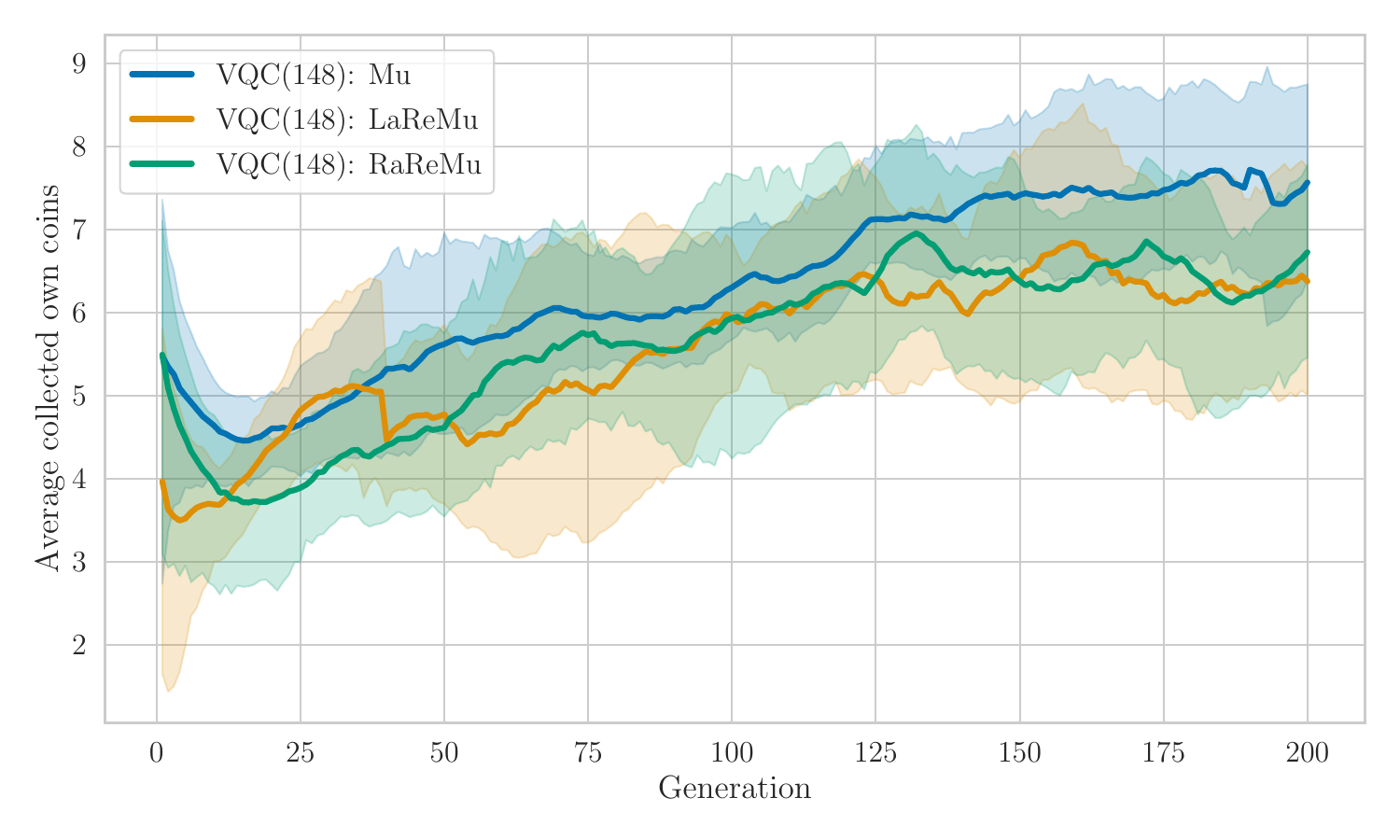}\label{fig:MethodsAvgownCoins}}
     \subfloat[][Own coin rate]{\includegraphics[width=0.32\textwidth]{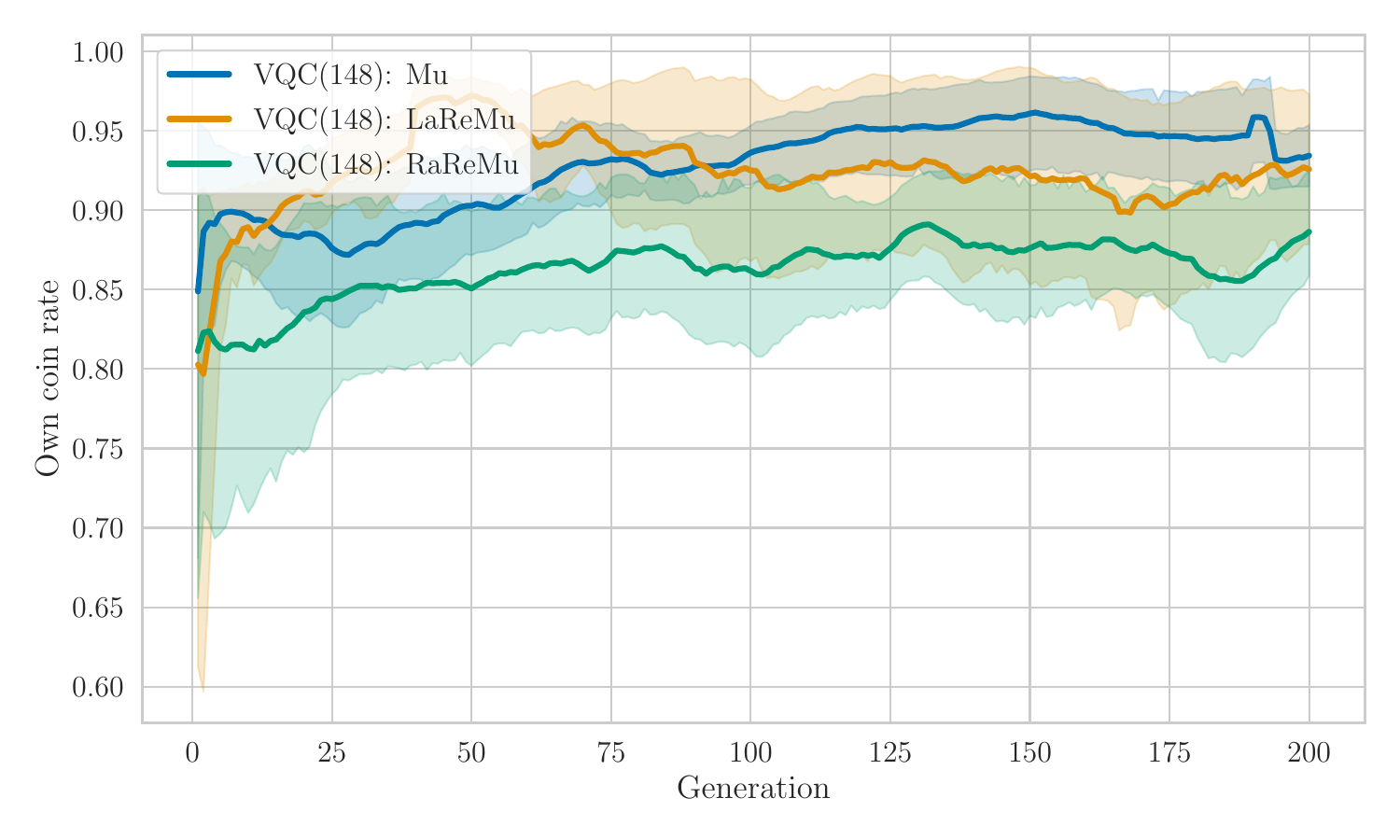}\label{fig:Methodsowncoinrate}}
     \caption{Comparison of (a) average coins collected, (b) average own coins collected and the own coin rate (c) in a 50 step Coin Game each generation, averaged over 10 seeds.}
     \label{fig:Methodscoins_comp}
\end{figure*}

\subsection{Comparing Generational Evolution Strategies}
\label{sec: VQCxVQC}

We aim to understand the impact of different generational evolution strategies. In this section, we contrast the performance of a mutation-only strategy (Mu) against two combined strategies of mutation and recombination. The first combined approach involves a crossover recombination strategy at a randomly chosen point in the parameter vector (RaReMu), while the second employs a layerwise crossover (LaReMu). Here, we choose a random layer and apply the crossover after the last parameter of the selected layer in the parameter vector. For all strategies, the mutation power $\sigma$ is fixed at $0.01$.

Examining the average scores depicted in \cref{fig:MethodsAvgScore}, the mutation-only strategy emerges as the best strategy. The crossover strategies showcase similar performance, though the layerwise method frequently achieves marginally superior outcomes. The mutation-only strategy starts with an average reward of $5$, dips slightly below $4$ by the 17th generation, and then steadily rises until the 140th generation. From this point, it fluctuates around a score of $7$. In contrast, the layerwise recombination begins at a lower $3.3$, experiences a rapid ascent until the 30th generation, then stabilizes, eventually reaching an average reward of $6$ by the 123rd generation. This is followed by pronounced fluctuations around this value. The random crossover strategy starts close to the mutation-only at $4.7$, but quickly descends to $3$ by the 17th generation. It then steadily climbs until the 131st generation, achieving a score of $6$. However, this score is not sustained and eventually settles around $5.5$, making it the least effective of the three methods.

Beyond score comparison, we evaluated the average number of coins collected during the experiments. As inferred from the scores, the mutation-only strategy consistently collects more coins, as shown in \cref{fig:MethodsAvgCoins}. Although there are periods where the strategies yield almost identical coin counts, at other times, a gap of up to $2$ coins is evident. On average, the combined strategies lag slightly behind the mutation-only in terms of coin collection.

The layerwise crossover's initial surge in \cref{fig:MethodsAvgScore} correlates with the uptrend in collected coins shown in \cref{fig:MethodsAvgCoins} and the coin rate detailed in \cref{fig:Methodsowncoinrate}. When comparing the strategies based on these metrics, mutation-only consistently achieves the highest coin rate over all generations and collects the most coins, accounting for its superior average reward. The random crossover strategy, while collecting more coins than the layerwise approach, has a significantly reduced coin rate, resulting in diminished overall rewards.

Exploring the coin rate, depicted in \cref{fig:Methodsowncoinrate}, the layerwise strategy leads until the 90th generation. After that, its rate declines, while the mutation-only strategy exhibits a gradual, consistent rise. A higher coin rate indicates enhanced agent cooperation within the testing environment. This coin rate, combined with the number of coins collected, determines an agent's reward.

In summary, the mutation-only strategy outperforms the combined strategies in our experiments. It not only garners the highest reward but also aligns best with our objective: maximizing reward. Hence, subsequent experiments will exclusively employ the mutation-only approach for the VQCs.

\begin{figure}[ht]
    \centering
    \includegraphics[width=\linewidth]{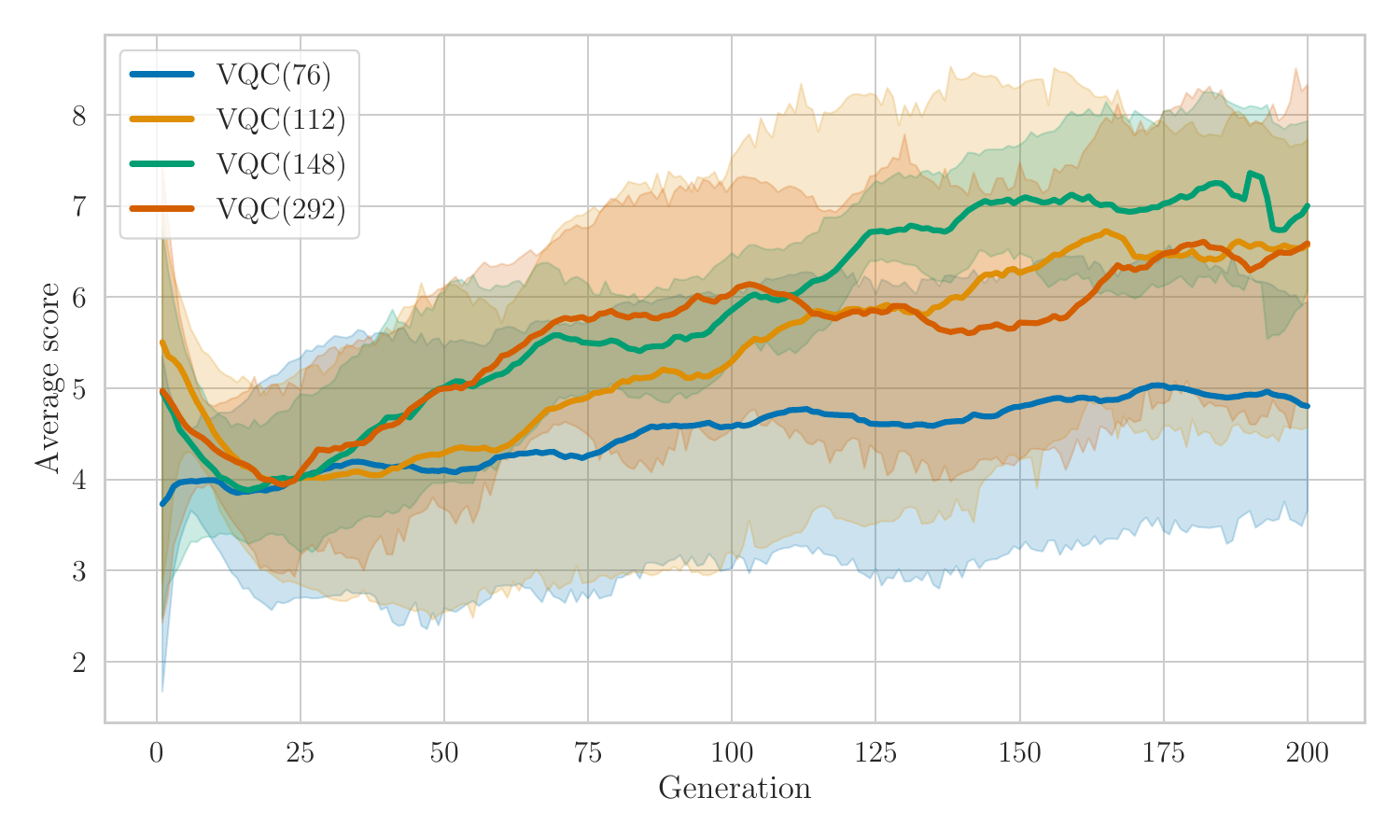}
    \caption{Average Score over the entire population. Each individual has completed 50 steps in the Coin Game environment each generation.}
    \label{fig:LayersAvgScore}
\end{figure}

\begin{figure*}
     \centering
     \subfloat[][Total coins collected]{\includegraphics[width=0.32\textwidth]{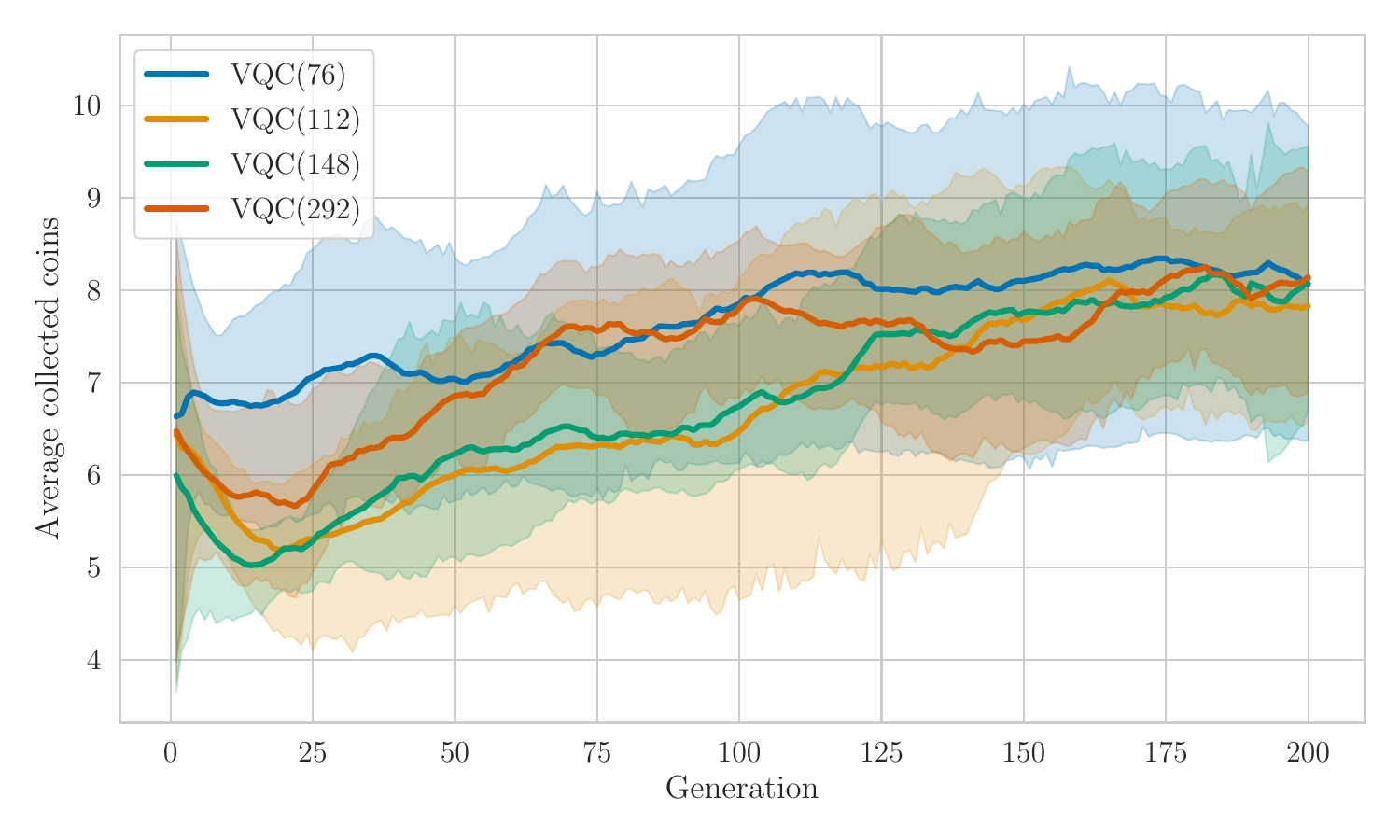}\label{fig:LayersAvgCoins}}
     \subfloat[][Own coins collected]{\includegraphics[width=0.32\textwidth]{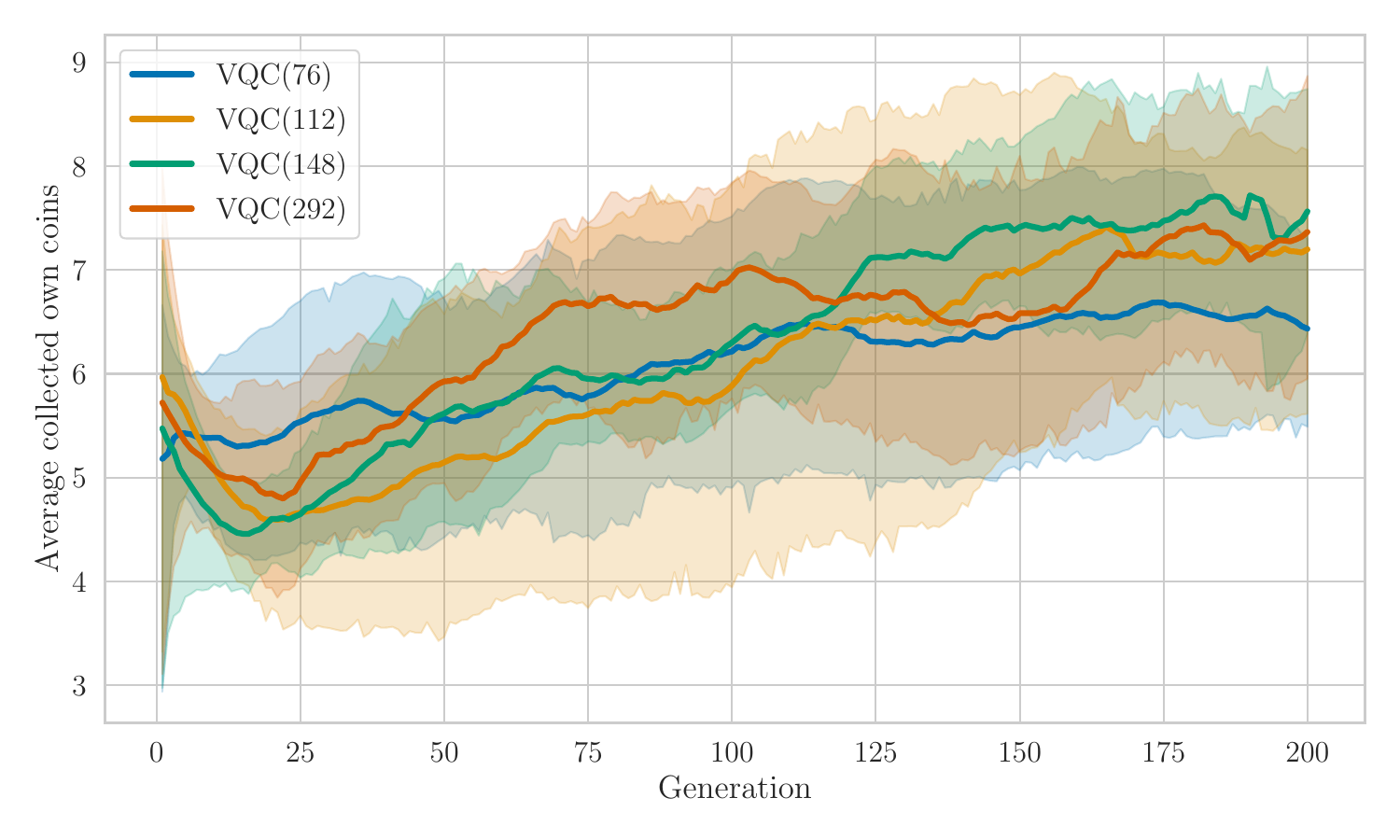}\label{fig:LayersAvgownCoins}}
     \subfloat[][Own coin rate]{\includegraphics[width=0.32\textwidth]{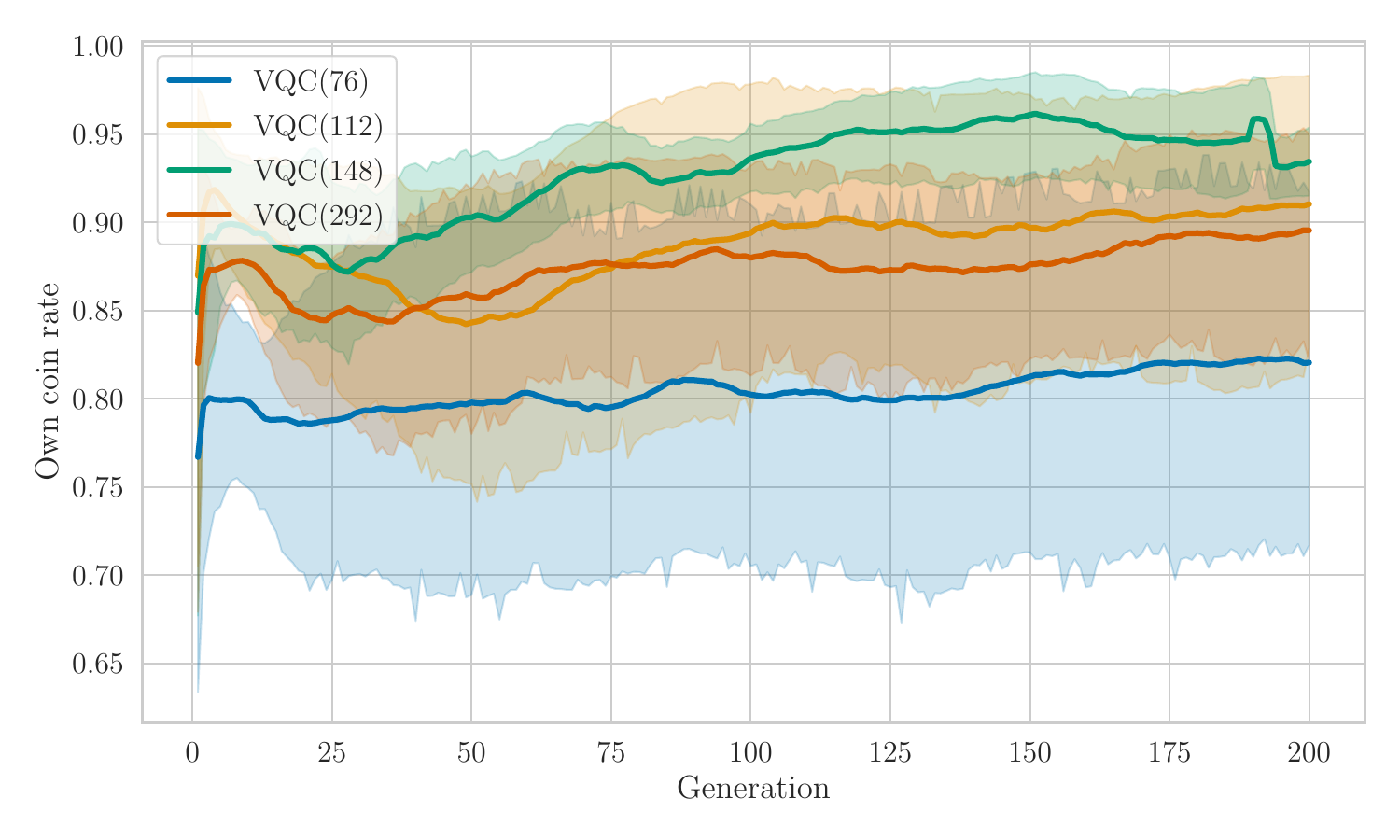}\label{fig:Layersowncoinrate}}
     \caption{Comparison of (a) average coins collected, (b) average own coins collected and the own coin rate (c) in a 50 step Coin Game each generation, averaged over 10 seeds.}
     \label{fig:Layercoins_comp}
\end{figure*}

\subsection{Assessing Varying Layer Counts}
\label{sec: VQCLayers}

We investigate the performance dynamics of VQCs with different layer counts, specifically with 4, 6, 8, and 16 layers. The relationship between layer counts and parameters is governed by the formula $3 * n * 6 + 4$, where $n$ stands for the number of layers. Accordingly, VQCs with 4, 6, 8, and 16 layers utilize 76, 112, 148, and 292 parameters respectively. We trained all VQCs using the mutation-only approach, setting the mutation strength to $\sigma = 0.01$.

Inspecting the average rewards in \cref{fig:LayersAvgScore}, all VQCs, bar the 4-layered one which starts slightly below $3$, commence with scores ranging from $5$ to $5.5$. By the 25th generation, each VQC stabilizes around a reward of $4$. The 4-layer VQC then gradually ascends, consistently holding an average reward of $5$ from the 175th generation. The 6-layer VQC exhibits a steady rise until the 62nd generation, with a more pronounced increase after that, peaking at $6.7$ around the 165th generation and subsequently oscillating around $6.5$. The 8-layer VQC consistently outperforms the others, reaching a reward of $5.5$ by the 70th generation, encountering a brief plateau, and then climbing to $7$ by the 140th generation. The 16-layer VQC, meanwhile, showcases a pronounced growth phase between the 25th and 70th generations, stabilizing around $6$ before another rise to $6.5$ around the 160th generation.

For a comprehensive understanding, we next probe the average coin collection in \cref{fig:LayersAvgCoins}. The 4-layer VQC consistently tops the coin collection metric, progressing from just below $7$ to $8$. The 6-layer VQC commences at $6.5$, dips to $5.2$ by the 23rd generation, and then rises to 8 by the 165th generation. The 8-layer VQC, despite securing the highest average reward, begins at $6$ and only stabilizes around $8$ after the 180th generation. The 16-layer VQC, after an initial dip, witnesses a rapid increase from the 24th to 103rd generation, briefly declines, and then fluctuates around $8$ coins. Towards the concluding generations, VQCs with more than 4 layers converge to collect approximately 8 coins.

Analyzing the own coin count in \cref{fig:LayersAvgownCoins}, we observe that, except for the 4-layer VQC, all VQCs initially decline before ascending. The 4-layer VQC displays a steady yet modest climb, concluding at a count of $6.5$. The 6-layer VQC takes the longest to commence its ascent, eventually oscillating around a count of $7.2$. The 8-layer VQC initiates its climb earlier, achieving a slightly higher count of $7.5$ by the end. The 16-layer VQC, notable for its rÏapid early ascent, consistently hovers around a count of $7$ after the 100th generation. Among the VQCs, the 4-layer variant lags, collecting over one own coin fewer than its counterparts.

Focusing on the own coin rate, the 4-layer VQC performs the worst. The performance parallels between the 6-layer and 16-layer VQCs are evident, both in terms of own coin rate and overall reward. The standout remains the 8-layer VQC, which, with its superior own coin rate and comparable coin count, has the highest reward.

In conclusion, our tests spotlight the 8-layer VQC as the top performer. Consequently, we select it combined with the the optimal evolutionary strategy outlined in \ref{sec: VQCxVQC}, for all further experiments. This section underscores that a higher layer count doesn't guarantee superior performance – the 16-layer VQC falls short of the 8-layer VQC's achievements. The experiments, however, don't conclusively establish the performance dynamics beyond 200 generations.

\begin{figure}[ht]
    \centering
    \includegraphics[width=\linewidth]{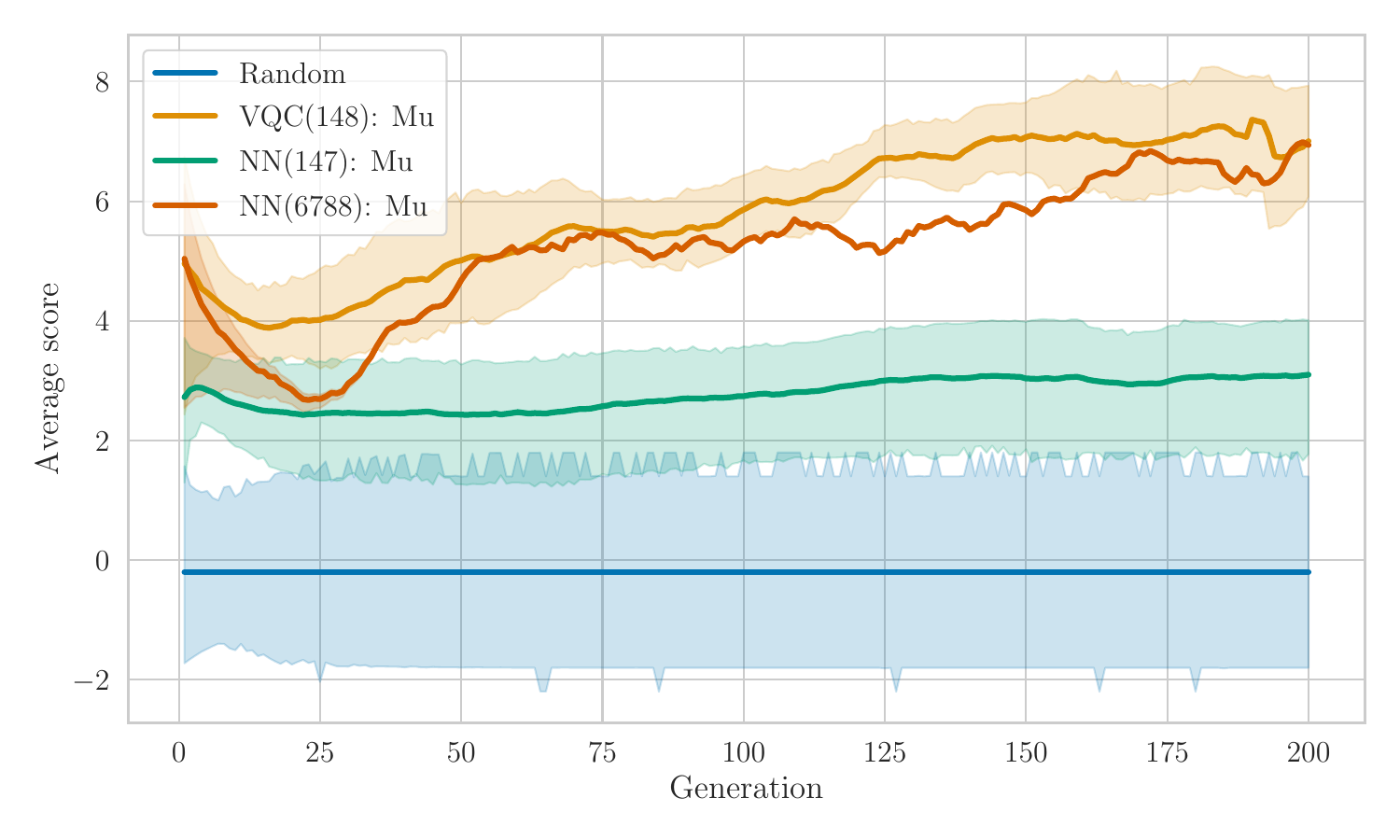}
    \caption{Average Score over the entire population. Each individual has completed 50 steps in the Coin Game environment each generation.}
    \label{fig:BestAvgScore}
\end{figure}

\begin{figure*}
     \centering
     \subfloat[][Total coins collected]{\includegraphics[width=0.32\textwidth]{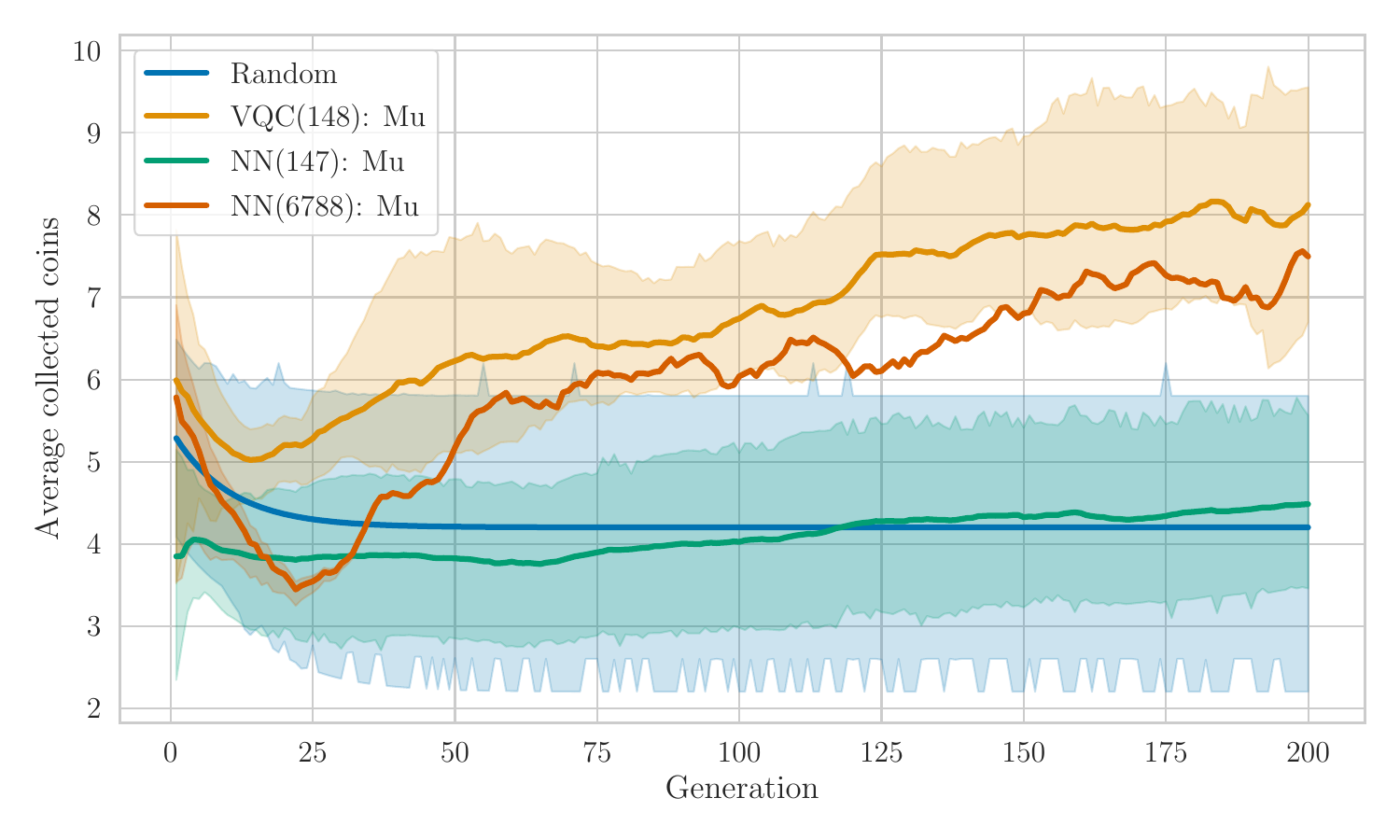}\label{fig:BestAvgCoins}}
     \subfloat[][Own coins collected]{\includegraphics[width=0.32\textwidth]{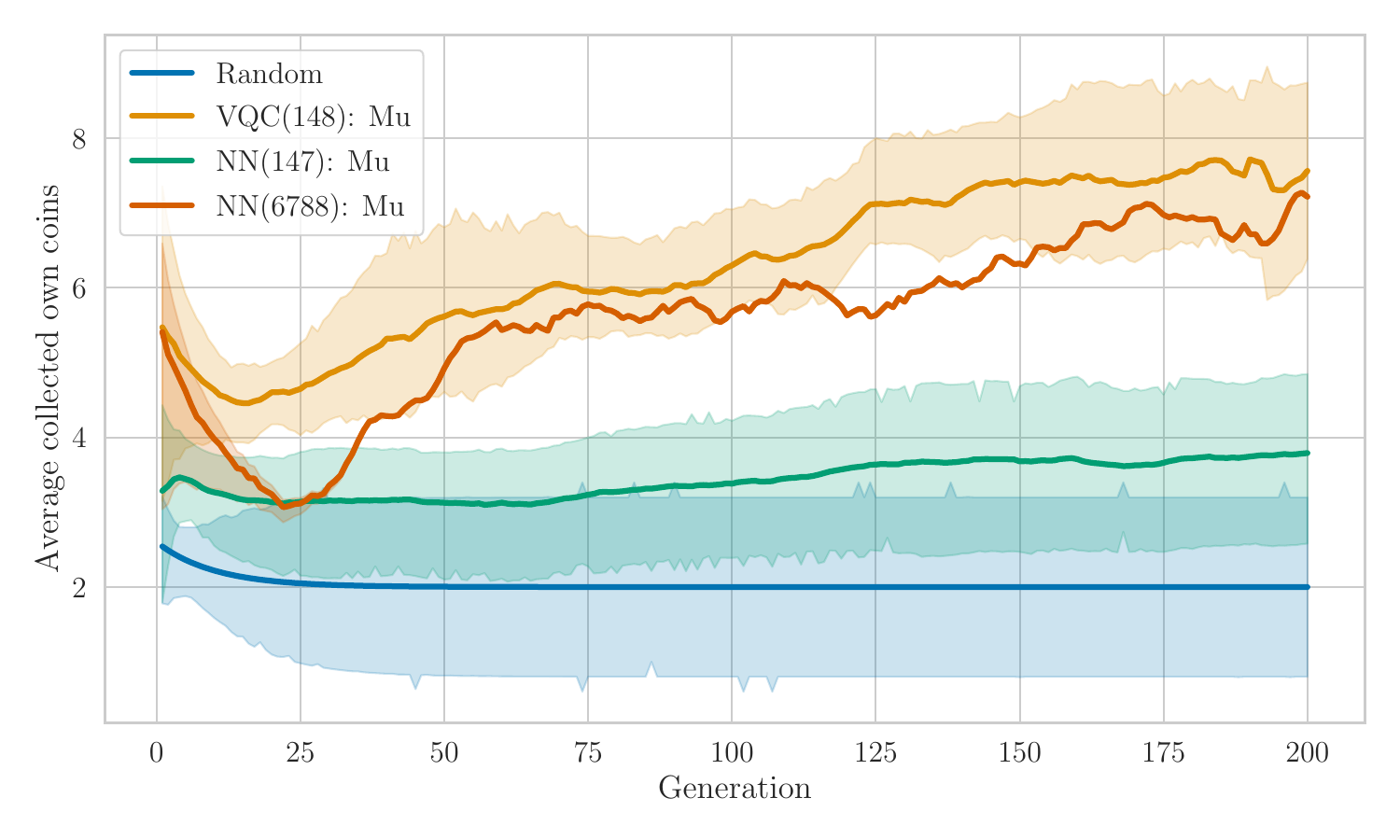}\label{fig:BestAvgownCoins}}
     \subfloat[][Own coin rate]{\includegraphics[width=0.32\textwidth]{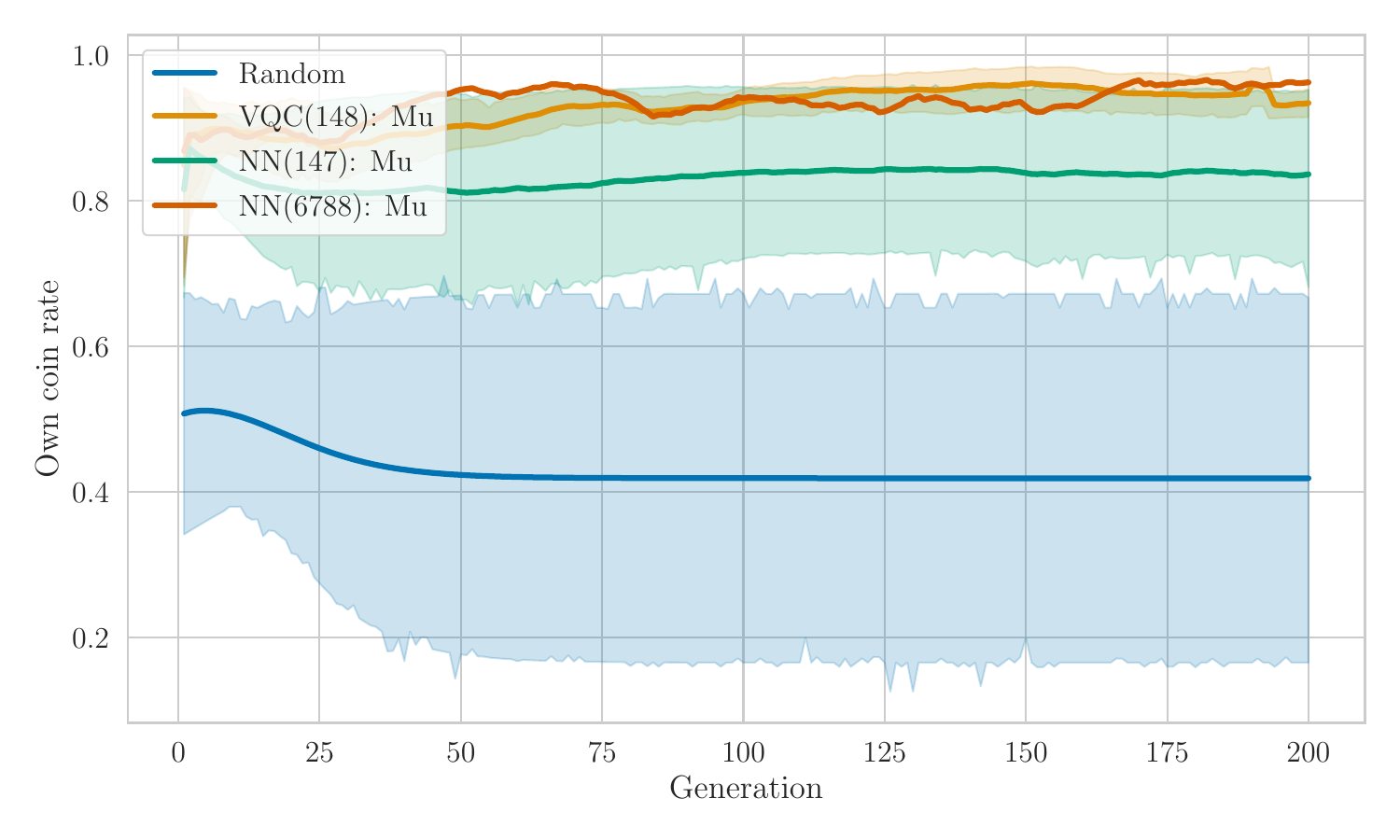}\label{fig:Bestowncoinrate}}
     \caption{Comparison of (a) average coins collected, (b) average own coins collected and the own coin rate (c) in a 50 step Coin Game each generation, averaged over 10 seeds.}
     \label{fig:Bestcoins_comp}
\end{figure*}

\subsection{Comparing Quantum and Classical Approaches}
\label{sec: VQCvs}
\subsubsection{Comparing VQC and Random}
\label{sec: VQCxRandom}
First, we compare the results of our VQC approaches to the results of the random baseline. 
In \cref{fig:BestAvgScore}, the score for the random acting agents is approximately $0$, since the cooperative sequential coin game is a zero-sum game. The evolutionary-trained VQC approache, however, perform significantly better leading to an average score around $7$. The total coins collected depicted in \cref{fig:BestAvgCoins} suggests that in contrast to the random agents, the VQC agents successfully learn to collect coins. The own coins collected correlates with number of collected coins (\cref{fig:BestAvgownCoins}). In \cref{fig:Bestowncoinrate} we can see that in neither case the cooperation increases over time. In summary, trained agents performed significantly better than random on all metrics, indicating that the training was successful.

\subsubsection{Comparing VQC and Small NN}
\label{sec:VQCxsmallNN}
As depicted in \cref{fig:BestAvgScore}, a better result is achieved by the VQC approach compared to random. On this basis, we compare the performance of this VQC approach to that of a neural network with a comparable number of parameters.
Here we exploit the higher expressive power of VQCs compared to conventional neural networks \cite{chen2022variational}. Similar to \cite{chen2022variational}, we define the expressive power as the capacity to represent particular functions with a constrained number of parameters. Note that the VQC has $148$ parameters (3 * 6 * 8 + 4). The neural network uses two hidden layers with dimension 3 and 4 respectively, resulting in a parameter count of $147$. Both the neural network and the VQC, are trained with mutation only with mutation power $\sigma = 0.01$. In \cref{fig:BestAvgScore}, we can see that the neural network reward fluctuates in the range of $2.5$ to $3$. As previously discussed in the last section, the VQC approach exhibits a slow learning curve leading to a significant higher score therefore consequently outperforming this neural network. The inferior performance can be explained by the small number of hidden units and parameters present in neural networks. Typically, the number of hidden units is chosen much higher. Further evidence of the neural network's deficiency is provided by the average number of coins collected. As shown in \cref{fig:BestAvgCoins}, the NN's number of collected coins is below the average score of the random agents until generation $115$ and after that slightly over it. In comparison, the VQC with the same number of parameters collects two times as many coins on average. The neural network is able to outperform random agents on the basis of its collected own coins, what can be seen in \cref{fig:BestAvgownCoins}, leading to the better performance regarding the own coin rate (\cref{fig:Bestowncoinrate}). In terms of collected own coins and the own coin rate, the neural network performs significantly worse than the VQC with nearly the same number of parameters. A neural network with this few hidden units and, consequently, parameters is not able learn in the coin game environment successfully. This demonstrates the power of VQCs for RL archiving significantly higher with the same amount of parameters.

\subsubsection{ Comparing VQC and Big NN}
\label{sec: VQCvsbNN}
In the previous section, we observed that a neural network with the same number of parameters as the VQC of our approach cannot match the VQC's performance. We will now compare the results with a neural network that has significantly more parameters. Again, mutation only is used for the evolution of subsequent generations in both cases and the mutation power is $\sigma = 0.01$. We chose a fully connected NN with two hidden layers of size $64$, resulting in a parameter count of $6788$.
If we first examine the reward of the neural network and the VQC, we can see in \cref{fig:BestAvgScore} that both produce very similar results over time. Initially, the VQC has a slightly higher score. From generation $50$ onward, there are only minor differences in terms of average score. Overall, the two strategies yield a score of approximatly $7$. On the basis of our experiments, we can say that the VQC achieves nearly identical performance in the Coin Game environment compared to a neural network that has $46$ times more parameters.
In \cref{fig:BestAvgCoins}, the initial value of the VQC method is again higher than that of the neural network. Due to a steeper learning curve, the neural network is able to compensate the lower starting value and achieves only a slightly smaller number of collected coins. From there are no discernible differences between the two models. A similar performance of the two approaches can be seen in \cref{fig:BestAvgownCoins}, where the average number of own coins collected is shown. In Terms of own coin rate, at first, the VQC archives a slightly higher score. However, at Generation 25, the neural network initially achieves a better own coin rate before being slightly lower between Generation 80 and 162. In the end, the neural network is slightly better in terms of the own coin rate.\\
In summary, there is little difference between the outcomes of the two approaches, despite the neural network having $46$ times the number of parameters compared to the VQC. Thus, we can reduce the number of parameters in our experiments by $97.88\%$ without sacrificing performance using VQCs. Similar to results in \cite{chen2022variational}, the VQC exhibits a great expressive power in and we recommend it for future use in QRL.
\section[Conclusion]{\uppercase{Conclusion}} \label{sec:conclusion}
Gradient-based training methods are, as of time of writing, not suitable for MAQRL due to problems with barren plateaus and vanishing gradients \cite{instabilities,chen2022variational}.In this work, we presented an alternative approach to gradient based training methods for MAQRL. For our gradient-free approach used in the experiments, barren plateaus did not arise; however, it cannot be ruled out that such issues may occur in gradient-free methods under different conditions or in more complex scenarios \cite{larocca2024reviewbarrenplateausvariational}. We build our approach upon the evolutionary optimization process used by \cite{chen2022variational}, expanding it to Multi-Agent systems and different generational evolution strategies. We proposed three quantum approaches for MARL. All approaches use VQCs as replacement for neural networks. Two approach use recombination in addition to mutation and the third approach mutation only. For evaluation, we chose Coin Game as our testing environment due to it's cooperative setting and relatively small observation space. As baselines, we used random agents and classical agents with neural networks, which were also trained using the evolutionary algorithm. To achieve a fair comparison, we chose a neural network with a similar amount of parameters, as well as one with a hidden layer size of $64\times64$.

In our experiments, we showed that our VQC approach performs significantly better compared to a neural network with a similar amount of trainable parameters. Compared to the larger neural network, we can see that the VQC approach achieves similar results, showing the effectiveness of using VQCs in a MAQRL environment. We can reduce the number of parameters by $97.88\%$ using the VQC approach compared to the similarly good neural network. In comparison to previous works \cite{chen2022variational}, we used recombination in addition to mutation in our evolutionary algorithm, which performed worse than mutation alone in the tested setting. Additionally, we used more layers for the VQCs than previous works \cite{chen2022variational}, as they have yielded better results in the experiments.

In the future, the VQC results could be run not only on a quantum simulator, but on real quantum hardware to determine which and if there is a difference. Also, a comparison of the VQC approach with a gradient based neural network would be an option for future work. Another option would be to compare the VQC approach in terms of the number of parameters with a data reuploading method and see if this can solve the coin game similarly well with even fewer qubits. Additionally, we could work on the hyperparameters and see if even better results can be achieved by adapting them.

\section*{\uppercase{Acknowledgements}}
This work is part of the Munich Quantum Valley, which is supported by the Bavarian state government with funds from the Hightech Agenda Bayern Plus. This publication was created as part of the Q-Grid project (13N16179) under the ``quantum technologies - from basic research to market'' funding program, supported by the German Federal Ministry of Education and Research.

\bibliographystyle{apalike}
{\small
\bibliography{main}}


\end{document}